\documentclass[amsmath,amssymb,floatfix,prb,epsf,showpacs,twocolumn,graphicx]{revtex4}
\usepackage{amsmath,amssymb,natbib,bm,graphicx,url,epsfig}
\usepackage[ansinew]{inputenc}
\usepackage{bm}

\usepackage{graphicx}
\usepackage{epstopdf}
\usepackage{psfrag}
\usepackage{verbatim}
\usepackage{appendix}
\usepackage{float}

\newcommand{\be}{\begin{equation}}
\newcommand{\ee}{\end{equation}}
\newcommand{\bea}{\begin{eqnarray}}
\newcommand{\eea}{\end{eqnarray}}
\newcommand{\ud}{\mathrm{d}}
\newcommand{\bi}{\mathrm{i}}

\newcommand{\bkay}{\boldsymbol{k}}
\newcommand{\bomega}{\bar{\omega}_0}
\newcommand{\bwt}{\begin{widetext}}
\newcommand{\ewt}{\end{widetext}}

\begin{document}
\title{Self-consistent Eliashberg theory, $T_c$, and the gap function in electron-doped cuprates}
\author{Dhananjay Dhokarh and Andrey V. Chubukov}
\affiliation{Department of Physics, University of Wisconsin, Madison, Wisconsin 53706, USA}
\begin{abstract}
We consider normal state properties, the pairing instability temperature, and the structure of the pairing gap in electron-doped cuprates.  We assume that 
the pairing is mediated by collective spin excitations, with antiferromagnetism emerging with the appearance of hot spots. 
We use a low-energy spin-fermion model and Eliashberg theory up to two-loop order. 
We justify ignoring vertex corrections by extending the model to $N >>1$ fermionic flavors, with 
$1/N$ playing the role of a small Eliashberg parameter. 
We argue, however, that it is still necessary to solve coupled integral equations for the frequency dependent fermionic and bosonic self-energies, both in the normal and superconducting state. Using the solution of the coupled 
equations, we find an onset of  $d-$wave pairing at
 $T_c  \sim 30$ K,
 roughly three times larger than the one obtained previously [P. Krotkov and A. Chubukov, Phys. Rev. B 74, 014509 (2006)],
 where it was assumed that the equations for fermionic and bosonic self-energies decouple in the normal state. 
To obtain the momentum and frequency dependent $d$-wave 
superconducting gap, $\Delta ({\vec k}_F, \omega_n)$, we derive and solve the non-linear gap equation together with the modified  equation for the  bosonic self energy which below $T_c$ also depends on $\Delta ({\vec k}_F, \omega_n)$. 
We find that $\Delta ({\vec k}_F, \omega_n)$ is a non-monotonic function of momentum along the Fermi surface, with its node along the zone diagonal and its
maximum some distance away from it. We obtain $2\Delta_{\mathrm{max}} (T\rightarrow0) /T_c \sim 4$. 
We argue that the value of $T_c$, the non-monotonicity of the gap, and $2\Delta_{\textrm{max}}/T_c$ ratio are all in good agreement with the experimental data on electron-doped cuprates.
\end{abstract}

\maketitle

\section{Introduction}

The first high temperature superconductors, discovered by Bednorz and M\"uller in 1986 \cite{bmuller}, were hole-doped cuprates, with $\textrm{Sr}$
 doped into $\textrm{La}_2\textrm{CuO}_4$. 
 Electron-doped cuprates were subsequently discovered in 
1989, by doping $\textrm{Ce}$ into $\textrm{Nd}_2\textrm{CuO}_4$~\cite{takagi}. 
At present, there exist a variety of electron-doped cuprates 
 of the  form $\textrm{RE}_{2-x}\textrm{M}_x\textrm{CuO}_4$ where RE (rare earth lanthanides) is Nd or Pr and M is Ce or Th.  For a comprehensive review 
on electron-doped cuprates see Ref. \cite{armit}.   

\begin{figure}[!htp]
\begin{center}
\psfrag{k1}[][]{\small $k_x$}
\psfrag{k2}[][]{\small $k_y$}
\psfrag{Q}[][]{\small $\vec{Q}$}
\includegraphics[scale=0.6]{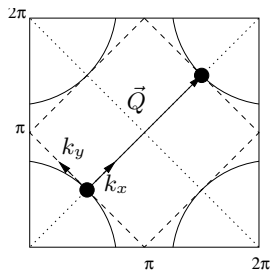}
\caption{\small Brillouin zone for electron-doped cuprates at the doping when hot spots merge on the zone diagonals. We argue that this doping is close to the one at which antiferromagnetic order emerges. The arcs represent the FSs, the dashed square represents the antiferromagnetic Brillouin zone. 
$k_x$ and $k_y$ are the directions of fermionic momenta transverse and along the FS, respectively.
\label{bqcp2}}
\end{center}
\end{figure}

Raman spectroscopy \cite{blumberg, qazilbash} and ARPES 
\cite{shen, ding, armita,matsui}  provide strong evidence that the symmetry of the superconducting gap in both electron- and hole-doped cuprates is 
$d_{{x}^2-{y}^2}$. This gap symmetry
 is also consistent with tunnelling \cite{fischer}, penetration depth \cite{hardy, kokales, proz, skinta}, and  Andreev reflections measurements \cite{deutscher}, although some of these experiments were originally interpreted differently. On the other hand,
maximum 
 $T_c$ in electron-doped cuprates is in the range of 
$10-30$ K, nearly an order of magnitude smaller than in hole-doped cuprates 
 despite the 
 ``Hubbard U'' being the same in both sets of materials (optical conductivity studies of parent hole- and electron-doped materials 
 reported 
 the same value, $1.7$ eV, of the optical gap \cite{dordevic, onose}).
 Additionally, the d-wave gap  in at least some electron-doped materials
 shows a non-monotonic behavior along the Fermi 
surface (FS),
 with the maximum 
 somewhere in between the zone diagonal (where the gap vanishes) and the region near $(0,\pi)$ (Fig.~\ref{bqcp2}).
 This non-monotonicity of the $d-$wave gap was 
predicted based on the analysis of Raman data~\cite{blumberg}, and was subsequently measured directly by ARPES in 
$\textrm{Pr}_{0.89}\textrm{LaCe}_{0.11}\textrm{CuO}_4$ (Ref. \cite{matsui}).  

One way to obtain $d_{x^2-y^2}$ pairing is to assume that the pairing is of electronic origin and is mediated by collective bosonic excitations at large momentum transfer, such that the interaction predominantly couples fermions with gaps of different signs (this effectively converts repulsive interaction into 
an attractive one). Both charge and spin fluctuations can give rise to such pairing.   
The idea of spin-fluctuation mediated pairing in  
cuprates has attracted substantial attention because antiferromagnetism is part of the phase diagram for both hole- and electron-doped cuprates. 
We follow earlier works on both hole and electron-doped cuprates~\cite{scal,pines,abanov, manske, tremblay, bansil, prelovsek} 
and other materials \cite{scaloh, bemonod,schmalian_organics}, and consider spin-mediated pairing.

In this paper we consider two issues. 
First, the origin of the smallness of  
 $T_c$ in electron-doped cuprates 
 compared to hole-doped cuprates; second, the momentum  and frequency dependence of the superconducting gap.

The relative smallness of $T_c$ in electron-doped cuprates 
is generally attributed to the fact that in these systems 
 $d-$wave pairing is less effective 
 because  hot spots (points at the FS separated by the antiferromagnetic momentum $\vec{Q}= (\pi,\pi)$)  are located much closer to the zone diagonals 
(see Fig. \ref{bqcp2}). 
This argument, however, is incomplete because magnetically-mediated $T_c$ 
remains finite and scales linearly with the overall spin-fermion coupling, even when hot spots merge at the zone diagonals \cite{chubukov}.
The complete picture is that in electron-doped cuprates, $T_c$ acquires the dependence of the FS curvature because  the velocities of hot 
fermions  become nearly antiparallel to each other when 
hot spots approach  zone diagonals (Fig. \ref{bqcp2}). Krotkov and one of us~\cite{chubukov} (hereafter referred to as KC) demonstrated that it is this dependence that gives rise to the additional numerical smallness of 
$T_c$. 
Using the same Eliashberg-type computational procedure as in hole-doped cuprates, KC obtained $T_c \sim 10$ K, using for the coupling the value extracted 
from the optical data at half-filling. 

Compared to experiment, the value of $10$ K
is a bit small,  particularly given that ``hot spot constrained'' calculations tend to somewhat overestimate $T_c$\cite{abanov_09}. 

In light of these points, we re-consider here the Eliashberg computational procedure. We argue that the earlier study neglected corrections of order one and should be therefore modified. To understand where the modification is required, we
 note that for superconductivity mediated by 
collective boson exchange 
Eliashberg-type
calculations of $T_c$ differ qualitatively from those of phonon-mediated superconductivity. For phonon superconductivity, the full phonon propagator is 
an input with the bosonic self-energy neglected because it is small by the same parameter as vertex corrections. For collective mode mediated pairing, bosonic dynamics originates from low-energy fermions and has to be computed together with the fermionic self-energy. As a consequence, one generally has to
solve a coupled set of integral equations for the fermionic self-energy, 
$\Sigma (\vec{k}, \omega)$, and the bosonic self-energy (the bosonic polarization operator), $\Pi (\vec{q}, \Omega)$.  

For hole-doped cuprates, the solution of this coupled set near hot spots 
is further simplified because bosonic self-energy, $\Pi(\vec{Q}, \Omega)$,  
has the form of Landau damping and does not depend on the fermionic self-energy. 
 As a result, the set of coupled equations for $\Sigma(\omega)\equiv\Sigma (\vec{k}_{\mathrm{hs}}, \omega)$ and $\Pi (\vec{Q}, \Omega)$ reduces to just one equation for 
$\Sigma (\omega)$, while $\Pi (\vec{Q}, \Omega)$ remains the same as for free fermions. 

The reasoning for the independence of $\Pi ({\vec Q}, \Omega)$ from $\Sigma (\omega)$  is the following: $\Pi ({\vec Q}, \Omega)$ is aconvolution of a particle and a hole, located near hot spots separated by $\vec{Q}$. When these two fermions have Fermi velocities directed at an arbitary angle w.r.t. each other, each fermionic line can be integrated over its dispersion $\epsilon_{\bkay}$, 
$\epsilon_{\bkay+\boldsymbol{Q}}$, and $\Pi ({\vec Q}, \Omega)$
emerges as the convolution of the two local propagators 
$\mathcal{G}_{\mathrm{l}} (\omega) = \int \ud \epsilon_{\bkay} \mathcal{G}(\vec{k}, \omega)$ and $\mathcal{G}_{\mathrm{l}} (\omega+\Omega) = \int \ud \epsilon_{\bkay+\boldsymbol{Q}} \mathcal{G}(\vec{k}+\vec{Q}, \omega+\Omega)$.  
The local 
$\mathcal{G}_{\mathrm{l}} (\omega) = -\bi \pi {\mathrm{sgn}} (\omega)$, and
  does not depend on fermionic $\Sigma (\omega)$; hence
 one does not need to dress up fermionic propagators.
An alternate way to state this is to observe that Landau damping is an anomaly, i.e., 
it can be re-expressed as the contribution from high-energies, where fermions are free particles.

Applying Eliashberg approach to electron-doped cuprates, KC 
assumed that the equations for $\Sigma$ and $\Pi$ decouple in this case too, so that they used 
the free-fermion  $\Pi(\vec{Q}, \omega)$ 
 to compute $\Sigma (\omega)$ and $T_c$. 
 However, $\Pi(\vec{Q}, \omega)$ for electron-doped cuprates scales as $\sqrt{|\Omega|}$ rather than as $|\Omega|$.
This is because 
 the Fermi velocities of hot
fermions 
 are anti-parallel, and the Jacobian of the transformation from $\int \ud^2\bkay$ to $\int \ud \epsilon_{\bkay} \ud \epsilon_{\bkay+\boldsymbol{Q}}$ diverges, so that $\Pi ({\vec Q}, \Omega)$ no longer reduces to the convolution of local propagators (alternatively, $\Pi ({\vec Q}, \Omega)$ is not an anomaly).  
As a consequence, $\Pi ({\vec Q}, \Omega)$ acquires the dependence on $\Sigma(\omega)$, and one has to solve the 
coupled set of equations for $\Sigma$ and $\Pi$.

We found that to a good numerical accuracy, the solution of the coupled set modifies previous results of KC by a constant factor $\alpha\sim 0.6$. 
That is, $\Pi (\vec{Q}, \Omega)$ by $\alpha$ and $\Sigma$ gets multiplied by  $1/\sqrt{\alpha}$.
Recalculating $T_c$ using these rescaled quantities, we find that 
$T_c$ increases by a factor $1/\alpha^2 \sim 3$. This brings Eliashberg $T_c$ to $30$ K, which is  quite consistent with the experimental data.   

This  $T_c \sim 30$ K is also in better agreement with $T_c$ 
 obtained within the FLEX approximation~\cite{manske}. In FLEX, full Green's functions are used,
but vertex corrections are neglected. The fact that for electron-doped cuprates we also need to use the full Green's functions 
makes Eliashberg theory 
more of a match with FLEX than 
for 
hole-doped cuprates.
Nevertheless, Eliashberg and FLEX are still not equivalent because in FLEX the full momentum dependent $\Sigma (\vec{k}, \omega)$ is used (and vertex corrections are neglected, though 
without any parametrical justification), while in our Eliashberg theory, only $\Sigma (\vec{k}_F, \omega)$ 
is relevant, while $\ud\Sigma (\vec{k}, \omega)/\ud\epsilon_{\bkay}$ and vertex corrections are neglected, by analogy with the electron-phonon problem.
 There is, however, no "natural" Eliashberg parameter analogous to $\omega_D/E_F$, and to justify the approximation we need to extend the theory 
to $N$ fermionic flavors and consider large $N$ limit; 
then $1/N$ becomes the Eliashberg parameter.

Very recent studies~\cite{max} have demonstrated that the large $N$ approximation in fact does not work beyond two-loop order. There are a series of multi-loop diagrams, beginning at three-loop order, which do not contain $1/N$, although apparently small numerically. We restrict to two-loop order, and do not address this issue in the present MS.           

The second issue we consider is the form of the superconducting gap as a function of momentum along the Fermi surface. KC obtained a non-monotonic gap 
function by solving for the $d$-wave eigenfunction of the gap equation at $T_c$. Here we solve the 
non-linear gap equation for $T<T_c$ 
(together with the coupled equation for the bosonic $\Pi ({\vec Q}, \Omega)$),  and find that the gap remains 
non-monotonic for $T<T_c$; there is a slight change in shape with $T$, but the non-monotonic character remains intact. At $T\rightarrow 0$, we find that the 
maximum value of the gap along the FS, $\Delta_{\mathrm{max}}$, is about two times larger than $T_c$, i.e., within Eliashberg theory, $2\Delta_{\mathrm{max}}/T_c \sim 4$. 
This ratio is quite consistent with these experimental data: from optics we have \cite{homes} $2\Delta_\mathrm{{max}}/T_c \sim 5$ 
for $\textrm{Pr}_{1.85}\textrm{Ce}_{0.15}\textrm{CuO}_4$, and from Raman measurements\cite{qazilbash} on $\textrm{Pr}_{2-x}\textrm{Ce}_x\textrm{CuO}_4$ and
$\textrm{Nd}_{2-x}\textrm{Ce}_x\textrm{CuO}_4$ we have $2\Delta_\mathrm{{max}} /T_c \sim 3.5$. 

The paper is organized as follows. In Sec. II we briefly review 
the spin-fermion model for electron-doped cuprates and its extention to $N >>1$.
In Sec. III we carry out normal state analysis. We show that vertex corrections are small by $1/N$, but that self-energy corrections to $\Pi(\vec{Q}, \Omega)$
are important. We then solve coupled equations for $\Sigma(\omega)$ and $\Pi(\vec{Q}, \Omega)$, and compare our results with those of KC, who used bare fermionic propagators to compute $\Pi(\vec{Q}, \Omega)$. We find that the fermionic self-energy scales as $(\bar{\omega}_0)^{1/4}|\omega|^{3/4}$, 
and we use this $\bar{\omega}_0$ scale as a measure for $T_c$.  
In Sec. IV we study the superconducting properties. 
We first solve for $T_c$, and then derive and solve the non-linear gap equation. The latter 
yields the non-monotonic $d-$wave gap 
$\Delta(\vec{k}_F, \omega_n)$,
which depends on momentum along the FS and also on Matsubara frequency.
Using this solution, we obtain
the ratio $2\Delta_{\mathrm{max}}/T_c$. In Sec. V we summarize and discuss our results. A vertex correction calculation is 
relegated to the Appendix.  

\section{The Model}

\begin{figure}[!htp]
\begin{center}
\includegraphics[scale=0.6]{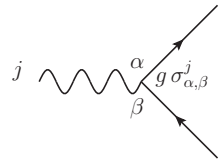}
\caption{The vertex for the spin-fermion interaction. 
The solid lines are fermions, the wavy line is a spin fluctuation.
\label{vrtx1}}
\end{center}
\end{figure}

Spin-fermion model has been discussed before, so we will be brief. The idea is that the low-energy physics of a system of itinerant fermions near antiferromagnetic instability, is adequately described by the interaction with low-energy, collective, bosonic excitations in the spin channel. The static part of a collective mode propagator comes from fermions at high energies and is the input for the low-energy theory, while the dynamical part comes from low-energy fermions and has to be calculated within the low-energy theory.
The Hamiltonian of the spin-fermion model is
\be\begin{split}\label{hamil}
\mathcal{H} &= \sum_{\bkay} \epsilon_{\bkay}c^{\dagger}_{\bkay, \alpha} c_{\bkay, \alpha} +\sum_{\boldsymbol{q}} \chi_{\mathrm{st}}^{-1} (\boldsymbol{q})
\boldsymbol{S}_{\boldsymbol{q}}\cdot \boldsymbol{S}_{-\boldsymbol{q}}\\
& + g\sum_{\boldsymbol{q}, \bkay, \alpha, \beta} c^{\dagger}_{\bkay+ \boldsymbol{q}, \alpha} \boldsymbol{\sigma}_{\alpha, \beta} c_{\bkay, \beta}\cdot \boldsymbol{S}_{-\boldsymbol{q}} 
\end{split}\ee
where $\epsilon_{\bkay}$ is the electronic dispersion, 
$c_{\bkay, \alpha}$ is the fermionic  operator for an electron with momentum $\bkay$ and spin $\alpha$, $\boldsymbol{\sigma}$ are the Pauli spin matrices,
 and $\boldsymbol{S}_{\boldsymbol{q}}$ is the bosonic variable
 describing collective spin degrees of freedom. Further, $g$ is the coupling constant and $\chi_{\mathrm{st}} (\boldsymbol{q})$ is the static spin susceptibility (the static propagator of collective spin degrees of freedom). The triple spin-fermion vertex is shown diagrammatically in Fig. \ref{vrtx1}. 
We will not show the spin indices explicitly in subsequent Feynman diagrams; it
should be understood that they are there, and proper accounting of them has been done to calculate the various prefactors while translating the 
diagrams into equations. 

For the static spin susceptibility, we follow earlier works and use the Ornstein-Zernike form
\be\label{chist}
\chi_{\mathrm{st}}(\vec{q}) = \frac{\chi_0}{\xi^{-2}+(q-Q)^2},
\ee  
where $\xi$ is the magnetic correlation length.  The constant $\chi$ should not be taken as another input because $g$ and $\chi_0$ 
only appear in the combination $\bar{g} = g^2\chi_0$. 
Throughout the paper we refer to  $\bar{g}$ as the effective spin-Fermion coupling constant.

Numerical studies of the onset of antiferromagnetic instability all show that antiferromagnetism emerges around the doping when hot spots merge along 
zone diagonals and the Fermi surface touches the antiferromagnetic Brillouin zone, as shown in Fig.~\ref{bqcp2}. The analytical argument is that the derivative of the static susceptibility of free fermions $\chi^0_{\mathrm{st}}(Q)$ with respect to doping 
diverges at this point, so that in its near vicinity $U \chi^0_{\mathrm{st}}(Q)$ reaches one and within RPA the full  $\chi_{\mathrm{st}}(Q) \propto 1/(1-U \chi^0_{\mathrm{st}} (Q))$ diverges. We assume for definiteness that 
the antiferromagnetic  quantum critical point (QCP), at which $\xi^{-1} =0$, is right at this doping.
Our results will change only a little if $\xi$ diverges not at this point, but close to it. 

As mentioned in the Introduction, the velocities of the fermions 
 along Brillouin zone diagonals 
 are anti-parallel to each other, with the
tangential component along the FS vanishing (see Fig. \ref{bqcp2}). The
dispersion of fermions near the hot spots is then 
 not the usual $\epsilon_{\bkay} = v_{Fx} k_x + v_{Fy} k_y$, but 
\be\label{spec}
\epsilon_{\bkay} = v_Fk_x+\beta^2 k_y^2
\ee
\be\label{spec1}
\epsilon_{\bkay+\boldsymbol{Q}} = -v_Fk_x+\beta^2 k_y^2
\ee
where $\beta^2$ parameterizes the curvature of the Fermi surface; this will play an important role in our analysis.   

It is convenient to measure the combined effect of the curvature and the interaction 
 in terms of the dimensionless parameter
\be\label{arr}
r = \frac{\bar{g}\beta^2}{\pi v_F^2}\ee
For parameters relevant to electron-doped cuprates, $r \sim 10^{-1}$. 
We also introduce for further use, the momentum scale 
 \be\label{kuzero} 
q_0 = \frac{\bar{g}}{\pi v_F}\ee 
and the frequency scale \be\label{megzero}
\omega_0 = \left(\frac{\bar{g}\beta}{2\pi v_F}\right)^2 =
 \left(\frac{q_0\beta}{2}\right)^2 =  \frac{\bar{g} r}{4\pi}
\ee 

The computational procedure is straightforward. We 
need to obtain the dynamic spin susceptibility, $\chi(\vec{q}, \Omega_n)$, in which the frequency dependence
comes through the total polarization operator $\Pi_{\mathrm{tot}}(\vec{q}, \Omega_n)$, which contains  contributions from both umklapp and non-umklapp processes. We have
\be\label{suscep}
\chi(\vec{q}, \Omega_n) = \frac{\chi_0}{(q-Q)^2 + \chi_0 \Pi_{\mathrm{tot}}(\vec{q},\Omega_n)}
\ee 
where
\be
\Omega_n= 2n\pi T
\ee is a Bosonic Matsubara frequency, and 
\be\Pi_{\mathrm{tot}}(\vec{q},\Omega_n) = 2(\Pi(\vec{q}, \Omega_n) +\Pi(-\vec{q}, \Omega_n))\ee

\begin{figure}[!htp]
\begin{center}
\includegraphics[scale=0.5]{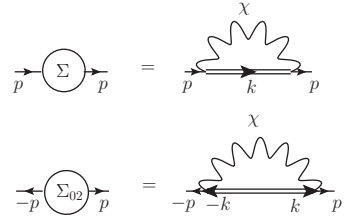}
\caption{Diagrams 
 for the normal and anomalous 
 fermionic self-energies ($\Sigma$  and $\Sigma_{02}$, respectively) 
in the Eliashberg-type theory (no vertex corrections).
The double lines are full Green's functions of intermediate fermions, 
with self-energies included.
\label{vself}}
\end{center}
\end{figure}

\begin{figure}[!htp]
\begin{center}
\includegraphics[scale=0.5]{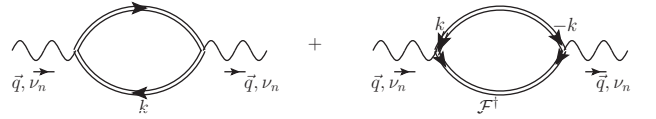}
\caption{Diagrams 
 for the polarization operator. The diagram on the right is 
 present only in the superconducting state.
 \label{pol1}}
\end{center}
\end{figure}

 We also 
 need to obtain 
 the full normal and anomalous Green's functions $\mathcal{G}(p)$
and $\mathcal{F}(p)$, respectively. These depend on the normal and anomalous self-energies, $\Sigma$ and $\Sigma_{02}$, as   
\be\begin{split}\label{gandf}
\mathcal{G} = \frac{X(-p)}{X(-p)X(p)+(\Sigma_{02} (p))^2}\\
\mathcal{F} = \frac{\Sigma_{02}(p)}{X(-p)X(p)+(\Sigma_{02}(p))^2}
\end{split}\ee
where $p = (\vec{p},\omega_n)$, $\omega_n = \pi T (2n+1)$ is a fermionic Matsubara frequency, and 
\be\label{xkay}
X(p) \equiv \mathcal{G}^{-1}_0(p) + \Sigma(p), ~~ \mathcal{G}^{-1}_0(p) = \bi\omega_n -\epsilon_{\boldsymbol{p}}.
\ee
In Eliashberg approximation,
 which we justify below, 
 the normal and anomalous fermionic self-energies are given by one-loop diagrams (Fig.~\ref{vself}) that generally 
involve the full fermionic 
propagators and the full dynamical spin susceptibility; similarly, the bosonic self-energy, $\Pi(\vec{q},\Omega_n)$, is the sum of bubbles made of 
the full fermionic propagators (Fig.~\ref{pol1}). We have,
\be\begin{split}\label{sigma}
\Sigma(p) = -3g^2T\sum_{m} \int\frac{\ud^2 \vec{k}}{(2\pi)^2} \mathcal{G}(\vec{k}, \omega_m)\chi(\vec{k}-\vec{p}, \omega_m-\omega_n)\\
\Sigma_{02}(p) = -3g^2T\sum_{m} \int\frac{\ud^2 \vec{k}}{(2\pi)^2} \mathcal{F}(\vec{k}, \omega_m)\chi(\vec{k}-\vec{p}, \omega_m-\omega_n)
\end{split}\ee
\be\begin{split}\label{mypi}
\Pi(\vec{q}, \Omega_n) &= 2g^2T\sum_{m} \int\frac{\ud^2 \vec{k}}{(2\pi)^2}\left(\mathcal{G}(\vec{k}, \omega_m)\mathcal{G}(\vec{k}+\vec{q}, \omega_m+\Omega_n)\right.\\
&\left.+ \mathcal{F}^{\dagger}(\vec{k}, \omega_m)\mathcal{F}(\vec{k}+\vec{q}, \omega_m+\Omega_n)\right)\end{split}\ee
These equations can be further simplified 
by integrating the r.h.s. over $\epsilon_{\bkay}$.  The dependence on 
$\epsilon_{\bkay}$ is in both fermionic and bosonic propagators, but the one in the bosonic propagator can be neglected because overdamped bosons are slow modes 
compared to fermions; 
 keeping the dependence on $\epsilon_{\bkay}$ in $\chi$,   
only gives rise to a small correction in the Eliashberg parameter. Within the
Eliashberg approximation, we also neglect the dependence of the fermionic self-energies on $\epsilon_{\bkay}$. With these steps the integration over $\epsilon_{\bkay}$ 
is straightforward.
For the normal state self-energy at a finite temperature we have
\be\begin{split}\label{sigmafTz2}
\Sigma(k_y, \omega_n) &= \frac{\bi 3g^2T}{4\pi v_F}\mathrm{sgn}(\omega_n)\times\\
&\sum_{|\Omega_m|<|\omega_n|}\int_{-\infty}^{\infty}\ud p_y\, \chi(k_y, p_y, \Omega_m)
\end{split}\ee  
 where 
 $k_y$ is the momentum component along the FS ($v_F k_x =- \beta^2 k^2_y$) and  $\chi(k_y, p_y, \Omega_m) = 
\chi(\vec{k}-\vec{p}, \Omega_m)|_{\epsilon_{\bkay} = \epsilon_{\boldsymbol{p}}=0}$ is the spin susceptibility at momenta connecting two points on the FS. Similarly, the linearized 
equation for the anomalous self-energy, valid at $T=T_c$, 
 is
\be\begin{split}\label{sigmaf3} 
\Sigma_{02}(k_y, \omega_n) &= - \frac{3g^2T}{4\pi v_F}\sum_{m} \int\ud p_y \frac{\chi(k_y, p_y, \omega_m-\omega_n)}{|\omega_m+\bi\Sigma(p_y,\omega_m)|}\times\\
&\Sigma_{02}(p_y, \omega_m)
\end{split}\ee 
We assume $d-$wave pairing for which $\Sigma_{02}(k_y, \omega_n) = - \Sigma_{02}(k_y + {\vec Q}, \omega_n)$. Shifting $p_y$ by ${\vec Q}$ in Eqs. (\ref{sigmafTz2}) and (\ref{sigmaf3}), we eliminate the overall minus sign on the r.h.s. of  (\ref{sigmaf3}). The polarization operator becomes, after the shift, 
$\Pi_{\mathrm{tot}} ({\vec k}-{\vec p}-{\vec Q}, \Omega)$ and can be approximated by
$\Pi_{\mathrm{tot}} ({\vec Q}, \Omega) = 4 \Pi  ({\vec Q}, \Omega)$. The susceptibility
spin susceptibility, $\chi(k_y, p_y, \Omega)$, then takes the form
\be\begin{split}
\label{dimlssuscepTz}
&\chi(k_y, p_y, \Omega_m) =\\
&\frac{\chi_0}{(k_y-p_y)^2+(\beta^2/v_F)^2 (k_y^2+p_y^2)^2+ 4\chi_0\Pi(\vec{Q}, \Omega_m)}\\
&= \frac{\chi_0}{q_0^2\left((k-p)^2+r^2(k^2+p^2)^2+\frac{4\chi_0}{q_0^2}\Pi(\vec{Q}, \Omega_m)\right)}
\end{split}\ee 

As mentioned in the Introduction, for fermion-fermion interaction mediated by a collective mode, there is no natural small Eliashberg-type 
parameter, analogous to $\omega_D/E_F$, that would make vertex corrections and $\ud\Sigma (k)/\ud\epsilon_k$ much smaller than one. To rigorously justify 
Eliashberg approximation, 
we need to extend the theory to $N>1$ fermionic flavors and take the $N >>1$ limit. Applying this extension to the 
low-energy model of Eq. \eqref{hamil}, we find that the polarization operator acquires an overall factor of $N$.  This factor also appears in the 
normal state fermionic self-energy, and can be absorbed into the renormalization of the frequency scale $\omega_0$ (given by Eq. \eqref{megzero}),
by redefining,
\be\label{omegaN}
\omega_0 \rightarrow \frac{\omega_0}{N^2}
\ee
Throughout the paper we will use this redefined $\omega_0$, keeping it fixed 
by rescaling the curvature $\beta$ by $N$.

\section{Normal state analysis}

We begin by citing the results from Ref.~\onlinecite{chubukov} for the normal state polarization operator and the normal state fermionic self-energy at the hot spot, obtained without self-consistency.
The polarization operator is given by 
\be\label{barepi}
\Pi_{0}({\vec Q}, \Omega_n) = \frac{N g^2}{2\pi\beta v_F}\sqrt{|\Omega_n|}
 F_\Pi\left(\frac{T}{|\Omega_n|}\right)
\ee where $F_\Pi (x\rightarrow 0) =1$, $F_\Pi (x>>1) \sim \sqrt{x}$, 
and the subscript $0$ is used to emphasize that $\Pi_0$ is obtained using bare fermionic propagators. The fermionic self-energy at the hot spot is given by 
\be\label{sigmaN}
\Sigma(\omega_n) = \bi (\omega_0)^{1/4}|\omega_n|^{3/4}\mathrm{sgn}(\omega_n) F_\Sigma \left(\frac{|\omega_n|}{\omega_0^{\mathrm{N}}}\right)
\ee 
where $F_\Sigma(x <<1) \propto x^{0.1}$ and $F_\Sigma (x >>1) \approx 1$.  
For the pairing problem we 
need the region  $|\omega_n| \geq \omega_0$, 
 for which $F_\Sigma$ can be approximated reasonably well by one.

We now discuss how these results are modified when we go beyond bare vertices and free fermion propagators. For definiteness, for the rest of this 
section we restrict our analysis to $T\rightarrow 0$. 

\subsection{Vertex corrections}

\begin{figure}[!htp]
\begin{center}
\includegraphics[scale=0.5]{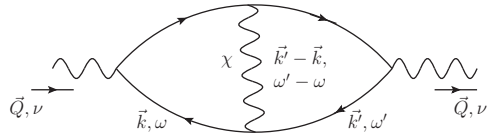}
\caption{
The polarization operator with the vertex correction, but without self-energy corrections to intermediate fermions.
\label{bvrtx}}
\end{center}
\end{figure}

We first show that diagrams with vertex corrections are small by $1/N$. As an example, consider the vertex correction diagram for the polarization 
operator (Fig.~\ref{bvrtx}).
We have 
\be\begin{split}\label{piv}
&\Pi_{\mathrm{v}}(\vec{Q},\Omega) = -2 N g^4\int_{-\infty}^{\infty} \frac{\ud\omega\ud^2 \bkay}{(2\pi)^3}\frac{\ud\omega'\ud^2\bkay'}{(2\pi)^3}
\mathcal{G}(\vec{k}, \omega)\times\\
&\mathcal{G}(\vec{k}+\vec{Q}, \omega+\Omega)\mathcal{G}(\vec{k}', \omega')\mathcal{G}(\vec{k}'+\vec{Q}, \omega'+\Omega)
\chi(\vec{k}-\vec{k}', \omega-\omega')
\end{split}\ee where the subscript v stands for $\Pi$ with vertex correction. 
The calculation is presented in Appendix A, and the result is 
\be
\Pi_{\mathrm{v}}(\vec{Q},\Omega) = \frac{g^2}{4\pi\beta v_F}\sqrt{\frac{\Omega}{8}}\log\left(4\sqrt{2}N^2\sqrt{\frac{\omega_0}{\Omega}}\right)
\label{a_4}
\ee
Comparing this with the result for $\Pi_0 (\vec{Q},\Omega)$ (Eqn \eqref{barepi}), we immediately see that 
 \be
\frac{\Pi_{\mathrm{v}}}{\Pi_0} = \frac{\log\left(4\sqrt{2}N^2\sqrt{\frac{\omega_0}{\Omega}}\right)}{4\sqrt{2} N},
\ee 
i.e., the vertex correction diagram for the polarization operator 
contains additional $\log N/N$ and can be safely neglected at large $N$.
One can verify that the same is true with the vertex correction diagram for the fermionic self-energy.

\subsection{Self-energy corrections to the polarization operator $\Pi$}

\begin{figure}[!htp]
\begin{center}
\includegraphics[scale=0.6]{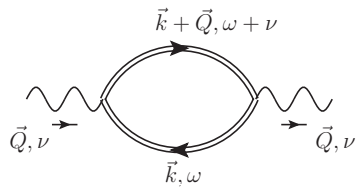}
\caption{The polarization operator without vertex corrections, but with full proparators of intermediate fermions.
\label{pol2}}
\end{center}
\end{figure}

\begin{figure}[!htp]
\begin{center}
\psfrag{nu}[][]{\small $x$}
\psfrag{rat}[tr][][1][-90]{\small $I(x)$}
\includegraphics[scale=.7]{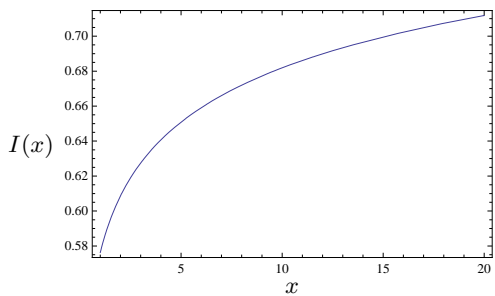}
\caption{Plot of the function $I(x)$ given by Eqn.~\eqref{rat1} vs. $x=\Omega/\omega_0$ for $x > 1$. Observe that $I(x)$ changes little between $x=1$ and $x \sim 10$. At very large $x$, $I(x)$ approaches one.
\label{bub4}}
\end{center}
\end{figure}

We next show that corrections from inserting  self-energy into the fermionic lines in the particle-hole bubble are not small in $1/N$ and should be included.
For this, we evaluate the polarization bubble $\Pi_{\mathrm{se}} (\vec{Q},\Omega)$ at $T\rightarrow 0$ with the renormalized Green's functions 
(Fig. \ref{pol2}) and compare the result with $\Pi_0 (\vec{Q},\Omega)$. We discuss full self-consistency later.  
The Green's function used is,  
\be
\mathcal{G}(\vec{k}, \omega_n) = \frac{1}{\bi\left(\omega_n + s(\omega_n) {\mathrm{sgn}}(\omega_n)\right) - \epsilon_{\bkay}}
\ee
where 
 \be
s(\omega_n) = (\omega_0)^{1/4}|\omega|^{3/4}
\ee
For the diagram for $\Pi_{\mathrm{se}}$ we have 
\bwt
\be
\Pi_{\mathrm{se}}(\vec{Q},\Omega) = 2Ng^2\int_{-\infty}^{\infty} \frac{\ud\omega\ud^2 \bkay}{(2\pi)^3}\frac{1}
{(\epsilon_{\bkay}-\bi(\omega+s(\omega)\mathrm{sgn}(\omega)))(\epsilon_{\bkay+\boldsymbol{Q}}-\bi(\omega+\Omega+s(\omega+\Omega)\mathrm{sgn}(\omega+\Omega)))}
\ee\ewt 
Expanding the dispersion around the hot spots, carrying out the integration over momenta, rescaling the resulting expression, and subtracting the 
 zero 
 frequency contribution (the static part is 
 already absorbed in $\chi^0_{\mathrm{st}}(\vec{Q})$),
we obtain  
\be\begin{split}
\Pi_{\mathrm{se}}(\vec{Q},\Omega) &= \frac{g^2 N}{2\pi \beta v_F} \sqrt{\Omega} I \left(\frac{\Omega}{\omega_0}\right)\\
&= \Pi_{0}(\vec{Q},\Omega)I \left(\frac{\Omega}{\omega_0}\right),
\end{split}\ee
where
\be\begin{split}\label{rat1}
I(x) &= -\frac{1}{\sqrt{x}}~\left(\int_{0}^{\infty}\frac{\ud y}{\sqrt{2y+x +y^{3/4}+(x+y)^{3/4}}}\right.\\ 
&\left. -\int_{0}^{\infty}\frac{\ud y}{\sqrt{2y +2 y^{3/4}}} \right) 
\end{split}\ee 
We plot $I(x)$ in Fig.~\ref{bub4}. 
This function monotonically increases with $x$ and approaches one at large $x$. This limiting behavior can be easily extracted from the 
integral for $I(x)$. 

We see that $\Pi_{\mathrm{se}}(\vec{Q},\Omega)$ is of the same order as $\Pi_{0}(\vec{Q},\Omega)$, i.e., there is no factor $1/N$ between them. This is not 
surprising as for typical $\Omega \sim \omega_0$ (i.e., for typical $x \sim O(1)$ in $I (x)$), $\omega$ and $\Sigma (\omega)$ are of the same order.  
This is in contrast to the vertex correction diagram in which the insertion of an additional bosonic propagator adds a $1/N$ factor. This factor could 
potentially be compensated by additional powers of momentum and frequency in the denominator; however it turns out that the integration over momentum 
and frequency in the diagram for $\Pi_\mathrm{v}$ only adds an additional $\log N$. 

We therefore conclude that
 it is possible to ignore vertex corrections; but 
 two-loop self-energy corrections to the polarization operator 
cannot be ignored. Therefore, we do need to calculate $\Sigma(\omega)$ and $\Pi(\vec{Q},\Omega)$ self-consistently. In this respect, the calculation of $\Sigma$ and $\Pi$ in electron-doped cuprates is qualitatively different from that in hole-doped cuprates, where $\Pi({\vec Q}, \Omega) \propto |\Omega|$ does 
not depend on $\Sigma$, i.e., the equations for $\Sigma$ and $\Pi$ decouple. 

We discuss our self-consistent solution in the next subsection.

\subsection{Self-consistent analysis of $\Sigma (\omega)$ and $\Pi(\vec{Q},\Omega)$}

For the self-consistent calculation of the bosonic self-energy, $\Pi$, and the normal state fermionic self-energy, $\Sigma(\omega)$, we use equations 
\eqref{sigmafTz2}, \eqref{dimlssuscepTz}, and the diagram in Fig. \ref{pol2}. In general, this would require one to solve the resulting coupled integral 
equations. In our case however, there is a way to simplify this calculation
because the function $I(x)$ in (\ref{rat1}) is quite flat 
at $x \geq 1$,  varying only by $8\%$ between  $x=1$
and $x=5$ (roughly from $0.6$ to $0.65$). To a reasonable accuracy, we can then approximate $I(x)$ by a constant $\alpha \sim 0.6$. 
Once we make this approximation, the polarization operator, $\Pi$, can be cast into the form
\be
\Pi_{\mathrm{se}} (\vec{Q},\Omega) = \frac{q^2_0}{4\chi_0} \sqrt{\Omega/\omega_0}
\ee
where
\be\label{bomegAl}
\bomega = \frac{\omega_0}{\alpha^2}
\ee
In other words, the dressed $\Pi_{\mathrm{se}}$ differs from $\Pi_0$ in Eq. (\ref{barepi}) by the replacement $\omega_0 \rightarrow \bomega$, without 
changing $q_0$. Because $\omega_0 = q^2_0 \beta^2/4$, this renormalization implies that the curvature $\beta$ gets effectively renormalized into 
${\bar \beta} = \beta/\alpha$.  
Substituting this polarization operator into equation \eqref{sigmafTz2} for the self-energy, we obtain, for  $\omega > \bomega$, 
\be\begin{split}
\Sigma(\omega) &= \bi \frac{3{\bar g}}{4\pi^2 v_Fq_0}\mathrm{sgn}(\omega)\int_{0}^{|\omega|}
\ud \Omega\int_{-\infty}^{\infty}\frac{\ud k}{\left(k^2+ \frac{4\chi_0}{q_0^2} \Pi_1(\vec{Q},\Omega)\right)}\\
&= \bi \frac{3}{4\pi}\mathrm{sgn}(\omega)\int_{0}^{|\omega|}
\ud \Omega\int_{-\infty}^{\infty}\frac{\ud k}{\left(k^2+\sqrt{\frac{\Omega}{\bomega}}\right)}\\
&= \bi (\bomega)^{1/4}|\omega|^{3/4}\mathrm{sgn}(\omega) 
\end{split}\ee 
Just as with $\Pi_{\mathrm{se}}$, the normal state
  fermionic self-energy preserves the form of Eq. \eqref{sigmaN}, with $\omega_0$ replaced by 
$\bomega$. Thus, by approximating $I (x)$ as a constant ($=\alpha)$, the full solution of the set of  self-consistent equations for $\Sigma$ 
and $\Pi (= \Pi_{\mathrm{se}})$ reduces to the replacement $\omega_0 \rightarrow \bomega$, with no change in $q_0$. 

\section{Superconducting properties}

\subsection{The value of  $T_c$}

The linearized gap equation (equation (\eqref{sigmaf3})  without self-consistent renormalization of $\Sigma$ and $\Pi$ 
(i.e., with $\omega_0$  instead of $\bomega$) has been solved by KC (Ref.\cite{chubukov}). These authors obtained
\be
T^{\textrm{KC}}_c = \omega_0 F (r) = \frac{r {\bar g}}{4\pi} F(r)
\ee
where, we remind, $r = \bar{g}\beta^2/(\pi v_F^2)$ is the dimensionless parameter proportional to the FS curvature. The function $F(r)$ is a 
non-monotonic function of $r$, with a flat maximum  near $r \sim 10^{-1}$.
Using ${\bar g} \sim 1.6$ eV (extracted from the charge transfer gap) and $t-t'$ dispersion 
with parameters taken from ARPES,
 KC obtained 
$\omega_0 \sim 10$ meV and $r \sim 0.08$, with $F(r) \sim 0.1$ (Ref. \cite{chubukov}). This yields $T^{\mathrm{KC}}_c \sim 10$ K. 

As we just discussed, the self-consistency of $\Sigma$ and $\Pi$ changes the scale $\omega_0$ into $\bomega = \omega_0/\alpha^2$, with 
 $\alpha\sim 0.6$. The  static part of the susceptibility in \eqref{dimlssuscepTz}, which provides the $r-$contribution in $F(r)$, is not affected by this renormalization, i.e., $F(r)$ does not change. 
 We then obtain,
\be\label{s_1}
T_c (r) = \bomega F(r) = \frac{T^{\mathrm{KC}}_c}{\alpha^2} \sim 3 T^{\mathrm{KC}}_c.
\ee so that $T_c$ changes by a factor of about $3$. For the same parameters as used by KC, we obtain $T_c \sim 30$ K.  
\begin{figure}[!htp]
\begin{center}
\psfrag{r}[][]{$r$}
\psfrag{Tc}[tr][][1][-90]{\small $T_c/\bar{g}$}
\includegraphics[width=\linewidth]{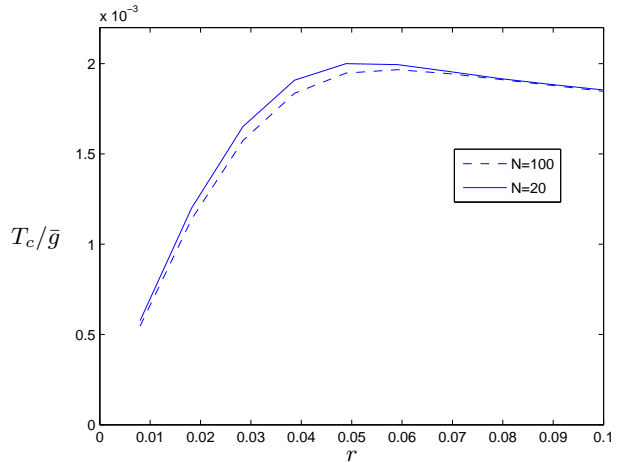}
\caption{$T_c$ in units of $\bar{g}$ versus r within the constant $\alpha$ approximation. The analytical formula is $T_c/{\bar g} = r F(r)/(4\pi \alpha^2)$, where $F(r)$ is a complicated function of $r$ (see Ref. \protect\cite{chubukov}).  The solid line is for $M=20$, 
the dashed for $M=100$, where $M$ is the number
of Matsubara points. The two curves are in good agreement, thus providing a check on our solution. We kept the number of points for the Gauss-Legendre (GL)
 quadrature fixed at $M_{\mathrm{GL}}=10$. 
\label{tcv}}
\end{center}
\end{figure}
\begin{figure}[!htp]
\begin{center}
\psfrag{k}[t][]{\small $\frac{k_y}{q_0}$}
\psfrag{S arb}[][][1][0]{\scriptsize $\Sigma_{02}\equiv \Delta$, arb. units}
\psfrag{T}[bl][][1][0]{\small $T=T_c$}
\includegraphics[scale=0.5]{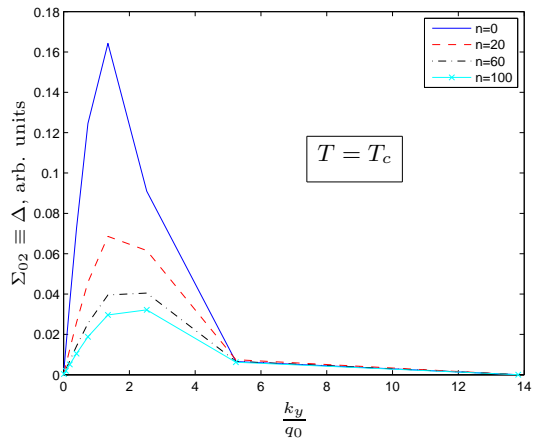}
\caption{\small $\Sigma_{02}(k_y, \omega_n)$ versus $k_y/q_0$, obtained as a normalized eigenvector of the linearized gap equation. We set $r = 0.08$ ($T_c \sim 0.30 \omega$). 
The magnitude of this $\Sigma_{02}$ does not have any significance as the actual $\Sigma_{02}$ is infinitesimally small at $T_c$.  
The $n$ in the legend correspond to different Matsubara frequencies, $\omega_n$.\label{idelta1}}
\end{center}
\end{figure}
For completeness we also 
computed $T_c(r)$ and the eigenfunction $\Delta (k_y, \omega_n)$ numerically, within our constant $\alpha$ approximation.  We used the 
Gauss-Legendre quadrature method, one of the many Gaussian type quadrature approximations \cite{reck}. 
We present the result for $T_c (r)$ in Fig. \ref{tcv}.
We find that the functional form of $T_c (r)$ is quite similar to the one obtained by KC, and the values are about $3$ times larger than theirs, as expected.
We also emphasize that $T_c (r)$ is rather flat over a wide range of $r \leq 10^{-1}$, so that any inaccuracy in determining the value of $r$ has little effect on $T_c$.

Our result is in much better agreement with experiments than $T^{\mathrm{KC}}_c$, particularly given that hot spot calculations tend to somewhat 
overestimate the value of $T_c$ compared to full scale calculation without expanding the fermionic dispersion near the hot spots~\cite{abanov_09}.

In Fig.~\ref{idelta1} we present the eigenfunction of the linearized gap equation for various Matsubara frequencies. It shows that the infinitesimally small $d-$wave gap at $T_c$ is non-monotonic along the FS, with its 
maximum not at the hot spot (the point $k_y=0$) but at a point away from it. This agrees with the result of KC. 

\subsection{The non-linear gap equation}

We next derive and solve the non-linear gap equation for $T<T_c$. The key goals here are to obtain the angular dependence of the gap along the FS, and 
to verify whether it remains non-monotonic below $T_c$. In the process, we also obtain the frequency dependence of the gap and compute the ratio 
$2\Delta_{\mathrm{max}}/T_c$.

The computational procedure is straighforward. We introduce $\Delta (\vec{k}, \omega_n)$ and fermionic $Z(\vec{k}, \omega_n)$ via
\be
\Sigma(\vec{k}, \omega_n) = -\bi\omega_n (1-Z(\vec{k}, \omega_n))
\ee \be
\Delta(\vec{k}, \omega_n) = \Sigma_{02}(\vec{k}, \omega_n)/Z(\vec{k}, \omega_n).
\ee 
The equations for the normal and anomalous, fermionic self-energies are re-expressed via $\Delta$ and $Z$ as \bwt
\begin{subequations}\be\label{z1}
Z(p) = 1 + \frac{3g^2 T}{\bi\omega_n}\sum_{m} \int\frac{\ud^2 \vec{k}}{(2\pi)^2}
\frac{\bi\omega_m Z(\vec{k}, \omega_m)+\epsilon_k}{(\omega_m^2+\Delta(\vec{k}, \omega_m)^2)Z(\vec{k}, \omega_m)^2+\epsilon_k^2 }\chi(\vec{k}-\vec{p}, \omega_m-\omega_n)\ee
\be\label{z2}\begin{split} \Sigma_{02}(p) &= \Delta(p)Z(p) = \\ & 3g^2T\sum_{m} \int\frac{\ud^2 \vec{k}}{(2\pi)^2}
\frac{\Delta(\vec{k}, \omega_m)Z(\vec{k}, \omega_m)}{(\omega_m^2+\Delta(\vec{k}, \omega_m)^2)Z(\vec{k}, \omega_m)^2+\epsilon_k^2}\chi(\vec{k}-\vec{p}, \omega_m-\omega_n)
\end{split}\ee\end{subequations}\ewt
where $p = ({\vec p}, \omega_n)$. 
We restrict momenta in $\Delta$ to the Fermi surface and introduce
\be
\left.\Delta(\vec{p}, \omega_m)\right|_{\epsilon_{\boldsymbol{p}}=0} = 
\Delta(p_y, \omega_m)
\ee
Carrying out the integration w.r.t. $\epsilon_{\bkay}$ as before, and using Eq. \eqref{z1} to eliminate $Z$ in the l.h.s. of Eq. \eqref{z2},
we obtain the non-linear gap equation in the form
\be\begin{split}\label{nonlingap1}
&\Delta({\tilde p}, \omega_n) =\\
&\frac{\frac{3T}{4{\bar \omega}_0^{\mathrm{N}}}\displaystyle{\sum_{m}} \int_{-\infty}^{\infty}\ud {\tilde k}\frac{\tilde{\chi}({\tilde p}, {\tilde k}, \omega_n-\omega_m)}{\sqrt{\omega_m^2+\Delta^2({\tilde k}, \omega_m)}}
\Delta({\tilde k}, \omega_m)}{1 + \frac{3T}{4{\bar \omega}_0^{\mathrm{N}}}\displaystyle{\sum_{m}} \int_{-\infty}^{\infty}\ud {\tilde k}
\frac{2m+1}{2n+1}\frac{\tilde{\chi}({\tilde p}, {\tilde k}, \omega_m-\omega_n)}{\sqrt{(\omega_m^2+\Delta^2({\tilde k}, \omega_m))}}}
\end{split}\ee
where we have  
\be\begin{split}\label{dimlssuscepT4}
&\tilde{\chi}(\tilde{p}, \tilde{k}, \Omega_n) =\\ 
&\frac{1}{\left((\tilde{p}-\tilde{k})^2 +r^2(\tilde{p}^2+\tilde{k}^{2})^2 +{\tilde \Pi}_\mathrm{tot}(\vec{Q}, \Omega_n)\right)}, 
\end{split}\ee and we introduced dimensionless 
$\tilde{k} = k_y/q_0$, $\tilde{p} = p_y/q_0$, and ${\tilde \Pi}_\mathrm{tot} = (\chi_0/q^2_0) \Pi_{tot} = (4\chi_0/q_0^2)\Pi$
The full polarization operator depends on $\Delta ({\tilde k}, \omega_n)$ and is given by
\be
\label{finTpisc2} \tilde{\Pi}_\mathrm{tot,sc}(\vec{Q}, \Omega_n) = -\frac{2T}{{\bar \omega}_0^{\mathrm{N}}}\sum_{m} \int_{-\infty}^{\infty}\ud {\tilde k}
(E_1 +E_2 + E_3)
\ee
where
\bwt
\be\label{e1e2e3}\begin{split}
E_1 &= -\bi\left(1+\frac{\omega_m}{\sqrt{\omega_m^2 + \Delta^2 ({\tilde k}, \omega_m)}}\right)\frac{8{\tilde k}^{2}+\bi\left((\omega_m+\Omega_n) - \sqrt{\omega_m^2+\Delta^2 ({\tilde k}, \omega_m)}\right)}
{(8{\tilde k}^{2}-\bi \sqrt{\omega_m^2+\Delta^2 ({\tilde k}, \omega_m)})^2+((\omega_m+\Omega_n)^2 +\Delta^2  ({\tilde k}, \omega_m + \Omega_n))}\\
E_2 &= \bi \left(1-\frac{\omega_m+\Omega_n}{\sqrt{(\omega_m+\Omega_n)^2+\Delta^2  ({\tilde k}, \omega_m + \Omega_n)}}\right)\frac{8{\tilde k}^{2} +\bi\left(\omega_m +\sqrt{(\omega_m+\Omega_n)^2+\Delta^2  ({\tilde k}, \omega_m + \Omega_n)}\right)}
{(8{\tilde k}^{2} + \bi \sqrt{(\omega_m+\Omega_n)^2+\Delta^2  ({\tilde k}, \omega_m + \Omega_n)})^2+ (\omega_m^2+\Delta^2  ({\tilde k}, \omega_m))}\\
E_3 &= \Delta  ({\tilde k}, \omega_m)\Delta  ({\tilde k}, \omega_m + \Omega_n)\left(\frac{1}{\sqrt{\omega_m^2+\Delta^2  ({\tilde k}, \omega_m)}
\left((8{\tilde k}^{2}-\bi \sqrt{\omega_m^2+\Delta^2  ({\tilde k}, \omega_m)})^2+((\omega_m+\Omega_n)^2 +\Delta^2  ({\tilde k}, \omega_m + \Omega_n))\right)}+\right.\\
&\left.\frac{1}
{\sqrt{(\omega_m+\Omega_n)^2+\Delta^2  ({\tilde k}, \omega_m + \Omega_n)}\left((8{\tilde k}^{2} + \bi \sqrt{(\omega_m+\Omega_n)^2+\Delta^2  ({\tilde k}, \omega_m + \Omega_n)})^2+(\omega_m^2+\Delta^2  ({\tilde k}, \omega_m))\right)}\right)
\end{split}
\ee\ewt
Eqs. \eqref{nonlingap1}) and \eqref{finTpisc2}) have to be solved self-consistently for $T<T_c$, to obtain the
superconducting gap $\Delta(\vec{k}_F, \omega_n)$ along the Fermi surface. 

\subsection{Numerical solution of the non-linear gap equation}

We  numerically solve the above set of equations using the Gauss-Legendre quadrature. We choose a particular value of 
$r\sim 0.08$ for which $T_c \sim 0.10\,\bomega \sim 0.30\,\omega_0$.    

\begin{figure*}
\begin{center}
\psfrag{k1}[t][]{\small $\frac{k_y}{q_0}$}
\psfrag{k2}[t][]{\small $\frac{k_y}{q_0}$}
\psfrag{k3}[t][]{\small $\frac{k_y}{q_0}$}
\psfrag{k4}[t][]{\small $\frac{k_y}{q_0}$}
\psfrag{D1}[tr][][1][-90]{\small $\frac{\Delta}{\omega_0}$} 
\psfrag{D2}[tr][][1][-90]{\small $\frac{\Delta}{\omega_0}$}
\psfrag{D3}[tr][][1][-90]{\small $\frac{\Delta}{\omega_0}$}
\psfrag{D4}[tr][][1][-90]{\small $\frac{\Delta}{\omega_0}$}
\psfrag{T1}[bl][][1][0]{\small $T=T_c$}
\psfrag{T2}[bl][][1][0]{\small $T=0.8T_c$}
\psfrag{T3}[bl][][1][0]{\small $T=0.5T_c$}
\psfrag{T4}[bl][][1][0]{\small $T=0.4T_c$}
\includegraphics[scale=0.5]{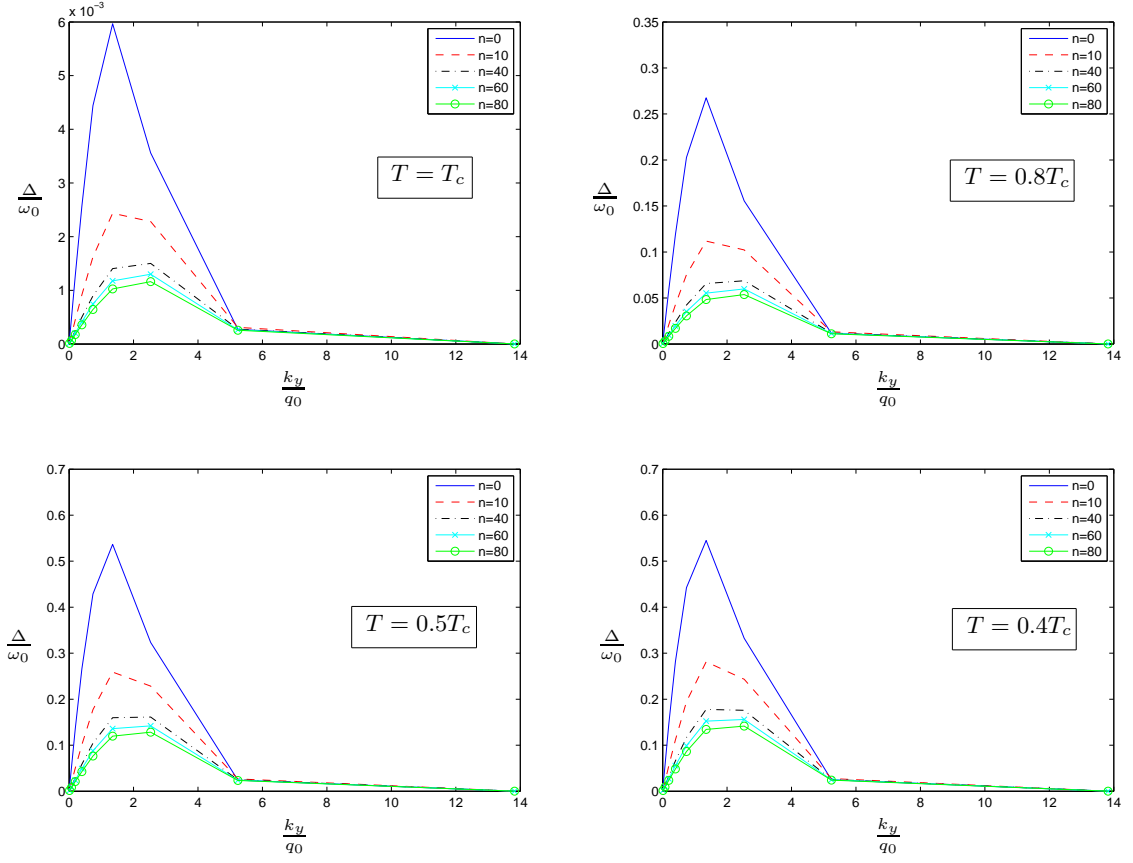}
\caption{\small The superconducting gap $\Delta(k_y, \omega_n)$ (in units of $\omega_0$) obtained as a solution of the non-linear gap
equation for different temperatures $T\leq Tc$. We set $r \sim 0.08$ ($T_c \sim 0.30\, \omega_0$) 
The index $n$ in the legend correspond to different Matsubara frequencies $\omega_n$.\label{odeltav}}
\end{center}
\end{figure*}

In Fig.~\ref{odeltav} we present our results for the solution of the coupled non-linear equations \eqref{nonlingap1} and \eqref{finTpisc2}, for various $T<T_c$. 
We see that the gap continues to be non-monotonic along the FS for all Matsubara frequencies, with its magnitude increasing as $T$ decreases, as indeed it should. The position of the maximum remains essentially intact. We consider our results as proof that the gap in electron-doped cuprates is indeed 
non-monotonic along the FS. We emphasize that the position of the maximum is \emph{not} at hot spots, which in our approximation are located right where
the zone diagonals intersect the FS. 

\begin{figure}[!htp]
\begin{center}
\psfrag{Dp}[tr][][1][-90]{\small $\frac{\Delta_\mathrm{max}}{T_c}$}
\psfrag{T}[t][]{\small $\frac{T}{T_c}$}
\includegraphics[scale=0.5]{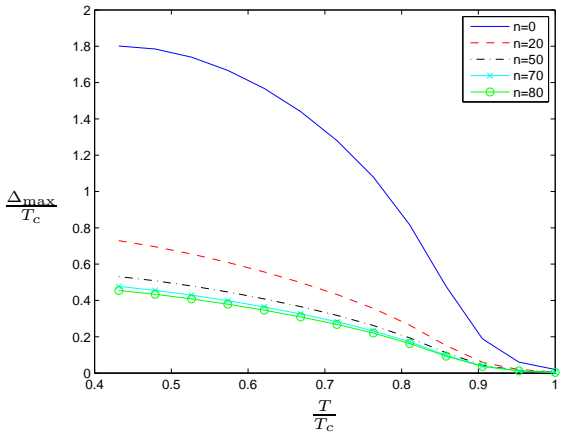}
\caption{\small The momentum peak of $\Delta(k_y, \omega_n)$ in units of $T_c$, plotted against $T/T_c$. Parameters are the same as in Fig. \protect\ref{odeltav}.
\label{deltapekV2}}
\end{center}
\end{figure}

In Fig.~\ref{deltapekV2}, we plot the temperature dependence of the gap maximum, at different Matsubara frequencies, measured in units of $T_c$.  We see that
for the lowest Matsubara frequency ($n=0$) the ratio \be
\frac{2\Delta_{\mathrm{max}}(T\rightarrow 0, n=0)}{T_c} \sim 4
\ee
The $n=0$ value of $\Delta(\vec{k}_F, \omega_n)$ is close to the real frequency $\omega = 0$ value of the gap; so we predict that the ratio of the measured
largest $2\Delta$ along the FS and $T_c$ should be close to four. 
This is not far from the BCS result for a $d$-wave superconductor~\cite{bcsdwave}. 
Experimentally, from optics we have \cite{homes} $2\Delta_\mathrm{{max}}/T_c \sim 5$
for $\textrm{Pr}_{1.85}\textrm{Ce}_{0.15}\textrm{CuO}_4$, and from Raman measurements\cite{qazilbash} on $\textrm{Pr}_{2-x}\textrm{Ce}_x\textrm{CuO}_4$ and
$\textrm{Nd}_{2-x}\textrm{Ce}_x\textrm{CuO}_4$ we have $2\Delta_\mathrm{{max}} /T_c \sim 3.5$.

\section{Summary discussion}

To summarize, we have considered in this paper normal state properties, the pairing instability temperature, and the structure of the pairing gap in 
electron-doped cuprates. We assumed that the pairing is mediated by low-energy, collective spin excitations, and that antiferromagnetic order develops close
to the doping where the FS touches the magnetic Brillouin zone boundary at four points, hot spots, which at this doping lie on the zone diagonals (i.e., 
antiferromagnetism emerges together with the appearance of hot spots). 

Because of the absence of a natural $\omega_D/E_F$ parameter, we extended the low-energy spin-fermion model, using Eliashberg theory, to $N >>1$ fermionic 
flavors with $1/N$ a small parameter. Using this we justified the neglect of vertex corrections and of 
$d \Sigma (k)/d\epsilon_k$; and also argued for the necessity to solve self-consistently, in the normal state, the coupled integral equations for fermionic 
$\Sigma (\omega)$ and bosonic $\Pi ({\vec Q}, \Omega)$. This is necessary because the self-energy corrections to the polarization bubble are not small
in $1/N$. In this respect, Eliashberg theory in electron-doped cuprates is different from the one in hole-doped cuprates. 
In hole-doped systems, the hot spots are far from the zone diagonals, Fermi velocities at hot spots separated by ${\vec Q}$ are directed at a finite angle
w.r.t. each other, and the polarization bubble, $\Pi ({\vec Q}, \Omega) \propto |\Omega|$,
is the convolution of local Green's functions which do not depend on the self energy $\Sigma(\omega)$.
In electron-doped cuprates, the hot spots at QCP are on the zone diogonals, the velocities of fermions separated by ${\vec Q}$ are antiparallel, and
the polarization bubble $\Pi ({\vec Q}, \Omega) \propto \sqrt{|\Omega|}$ is not expressed as a convolution of two local Green's functions. As a consequence, 
the self-energy corrections become relevant, necessitating a self-consistent analysis of $\Pi({\vec Q}, \Omega)$ and $\Sigma(\omega)$.

In the earlier study by KC~\cite{chubukov}, these self-energy corrections to the bubble were not included, and the $T_c$ calculation was carried
out without taking this self-consistency into account. In our work we have computed $\Sigma$ and $\Pi$ self-consistently. 
 We have argued that, to a reasonable accuracy, the self-consistent solution for $\Sigma$ and $\Pi$ reduces to the renormalization of earlier results by constant factors, i.e., $\Pi \rightarrow \alpha\Pi$ and $\Sigma \rightarrow \Sigma/\sqrt{\alpha}$, where $\alpha \sim 0.6$.   
 
Using these results, we re-derived the formula for $T_c$ and obtained $T_c \rightarrow T_c/\alpha^2 \sim 3T_c$. Using the same parameters as in the earlier study, we obtained $T_c \sim 30-35$K, in very good agreement with the experimental data for electron-doped cuprates. 

We then derived a coupled set of non-linear equations, for $T< T_c$, for the frequency and momentum dependent $d-$wave superconducting gap 
$\Delta ({\vec k}_F, \omega_n)$ along the FS and the polarization operator $\Pi ({\vec Q},\Omega_n)$. These equations are highly non-linear as $\Delta$ 
holds non-linear dependence on $\Pi$, and $\Pi$ a holds non-linear dependence on $\Delta$. Our numerical solution of this set of non-linear  
equations shows that $\Delta ({\vec k}_F, \omega_n)$ decreases monotonically, as expected, with increasing 
Matsubara frequency $\omega_n$; and more interestingly, that for each $\omega_n$ the gap is non-monotonic in momentum along the FS, with its node at the
zone diagonal (where a hot spot is also located), and the maximum some distance away from the zone diagonal. This non-monotonicity of the 
superconducting gap has been conjectured 
based on the analysis of Raman data \cite{blumberg} and subsequently detected in ARPES measurements.~\cite{matsui}   
 
For $2\Delta_{\mathrm{max}} (T\rightarrow 0, n=0) /T_c$ we obtained a value close to four. This is also consistent with the data: from optical measurements
$2\Delta_\mathrm{{max}}/T_c \sim 5$ (Ref.\cite{homes}), and from Raman measurements $2\Delta_\mathrm{{max}} /T_c \sim 3.5$ (Ref. \cite{qazilbash}).

\section{Acknowledgements}
We acknowledge helpful discussions with Mark Friesen, Ilya Eremin, James Rossmanith, and Tigran Sedrakyan. This work was supported by NSF-DMR-0906953.

\appendix 
\section{Calculation of $\Pi_{\mathrm{v}}$}
In this appendix we present the details of the calculation of the polarization operator with a vertex correction, Eq.~\eqref{piv}. 
We have 
\bwt
\be
\Pi_{\mathrm{v}}(\vec{Q},\Omega) = -2Ng^4\int_{-\infty}^{\infty} \frac{\ud\omega\ud^2 \bkay\ud\omega'\ud^2\bkay'}{(2\pi)^6}
\frac{\chi(\vec{k}-\vec{k'}, \omega-\omega')}{(\bi\omega-\epsilon_{\bkay})(\bi(\omega+\Omega)-\epsilon_{\bkay+\boldsymbol{Q}})
(\bi\omega'-\epsilon_{\bkay'})(\bi(\omega'+\Omega)-\epsilon_{\bkay'+\boldsymbol{Q}})}
\label{a_1}
\ee\ewt 
Using Eqs. \eqref{spec} and \eqref{spec1} for $\epsilon_{\bkay}$ and $\epsilon_{\bkay+\boldsymbol{Q}}$ and carrying out a change of variables, we re-write (\ref{a_1}) as
\bwt\be
\Pi_{\mathrm{v}}(\vec{Q},\Omega) = -\frac{2Ng^4}{v_F^2}\int_{-\infty}^{\infty} \frac{\ud\omega\ud\omega'\ud k_y\ud k'_y\ud\epsilon_{\bkay}\ud\epsilon_{\bkay'}\qquad\chi(\vec{k'}-\vec{k}, \omega'-\omega)}
{(\epsilon_{\bkay}-\bi\omega)(\epsilon_{\bkay}-2\beta^2 k_y^2+\bi(\omega+\Omega))(\epsilon_{\bkay'}-\bi\omega')(\epsilon_{\bkay'}-2\beta^2k'^2_y+\bi(\omega'+\Omega))}
\ee\ewt
Once again, neglecting the dependence of $\chi$ on $\epsilon_{\bkay}$, we set, 
\be\label{chif}
\chi(\vec{k'}-\vec{k}, \omega'-\omega) = \chi(\vec{k'}-\vec{k}, \omega'-\omega)|_{\epsilon_{\bkay} = \epsilon_{\bkay'}=0}
\ee Using equation \eqref{suscep} for $\chi$, and eliminating the dependence on $k_x$, $k'_x$ using 
\be\begin{split}
\epsilon_{\bkay} & = v_Fk_x+\beta^2k_y^2 = 0\\
\epsilon_{\bkay'} & = -v_Fk'_x+\beta^2k'^2_y = 0
\end{split}\ee we obtain, 
\be\begin{split}\label{dimlssuscep}
&\chi(k'_y, k_y, \omega'-\omega) = \chi(\vec{k'}-\vec{k}, \omega'-\omega)|_{\epsilon_{\bkay} = \epsilon_{\bkay'}=0}\\
&=\frac{\chi_0}{(k_y-k'_y)^2+\frac{\beta^4}{v_F^2}(k_y^2+k'^2_y)^2+4\chi_0\Pi(\vec{Q}, \omega'-\omega)}\\
&=\frac{\chi_0}{q_0^2\left((k-k')^2+r^2(k^2+k'^2)^2+N\sqrt{\frac{|\omega'-\omega|}{\omega_0}}\right)}.
\end{split}\ee 
In the last line we have replaced $k_y$ and $k'_y$ by dimensionless $k = k_y/q_0$ and $k' = k'_y/q_0$, and used Eq.~\eqref{barepi} for $\Pi$ with 
$F_\Pi =1$ because we are at  $T\rightarrow 0$. Also the $\omega_0$ that appears is as given by equation \eqref{megzero}, i.e. without the large $N$
redefinition. 
We have verified that for the present calculation the $r$ dependent term in $\chi$ is irrelevant and can be dropped.
The integrations w.r.t. $\epsilon_{\bkay}$, $\epsilon_{\bkay'}$ are factorized and easy to carry out.
The result is \bwt\be
\begin{split} \Pi_{\mathrm{v}}(\vec{Q},\Omega) = -\frac{2Ng^4}{v_F^2(2\pi)^6}\int\ud k_y\ud k'_y \left[2\pi\bi\int_{0}^{\infty}\frac{\ud\omega}{\bi(2\omega+\Omega)-2\beta^2k_y^2} + 
2\pi\bi\int_{-\infty}^{-\Omega}\frac{\ud\omega}{2\beta^2k_y^2-\bi(2\omega+\Omega)}\right]\\
\left[2\pi\bi\int_{0}^{\infty}\frac{\ud\omega'}{\bi(2\omega'+\Omega)-2\beta^2 k'^2_y} +
2\pi\bi\int_{-\infty}^{-\Omega}\frac{\ud\omega'}{2\beta^2k_y^2-\bi(2\omega'+\Omega)}\right]\chi(k'_y-k_y,\omega'-\omega)
\end{split}
\ee
Re-arranging the integrals over $\ud\omega$ and $\ud\omega'$,
we get a more compact result 
\be
\label{a_2}
\Pi_{\mathrm{v}}(\vec{Q},\Omega) = \frac{2Ng^4}{(2\pi)^4}\int\ud k_y\ud k'_y (\int_{0}^{\infty}\ud\omega\ud\omega'-
\int_{0}^{\infty}\ud\omega\int_{-\infty}^{-\Omega}\ud\omega')(D(\omega, k_y)D(\omega', k'_y) + \mathrm{c.c.})\chi(k'_y-k_y,\omega'-\omega)
\ee where \be
D(\omega, k_y) = \frac{1}{2\beta^2 k_y^2 - \bi(2\omega+\Omega)}
\ee
and 
\be
D(\omega, k_y)D(\omega', k'_y) + \mathrm{c.c.} = 2\frac{2\beta^2k^2_y2\beta^2k'^2_y-(2\omega+\Omega)(2\omega'+\Omega)}
{((2\beta^2k^2_y)^2+(2\omega+\Omega)^2)((2\beta^2k'^2_y)^2+(2\omega'+\Omega)^2)}
\ee
Substituting this into (\ref{a_2}) and using $2\beta^2 k^2_y = 8\omega_0 k^2$, where $k$ is the dimensionless variable introduced above, we obtain
\be\begin{split}\label{mypi1}
\Pi_{\mathrm{v}}(\vec{Q},\Omega) = &\frac{g^2}{2\pi\beta v_F}\frac{\sqrt{\omega_0}}{64\pi^2}(\int_{0}^{\infty}\ud(\omega/\omega_0)\ud(\omega'/\omega_0)-
\int_{0}^{\infty}\ud(\omega/\omega_0)\int_{-\infty}^{-\Omega}\ud(\omega'/\omega_0))\\
&\int\ud k\ud k'\frac{F_{\Omega}(k,k', \omega, \omega')}{(k-k')^2+N\sqrt{|\omega-\omega'|/\omega_0}}
\end{split}\ee
where \be
F_{\Omega}(k,k', \omega, \omega') = \frac{k^2k'^2-(\frac{2\omega+\Omega}{8\omega_0})(\frac{2\omega'+\Omega}{8\omega_0})}
{(k^4+(\frac{2\omega+\Omega}{8\omega_0})^2)(k'^4+(\frac{2\omega'+\Omega}{8\omega_0})^2)}
\ee

Extending this result to large $N$ as described in the main text (Eq. \eqref{omegaN}), i.e., redefining $\omega_0\rightarrow\omega_0/N^2$; and introducing new variables 
\be
x = \omega/\omega_0, ~~y = \omega'/\omega_0
\ee and
\be
\bar{k} = Nk,~~\bar{k'} = N k'
\ee 
we
obtain
\be\begin{split}
\Pi_{\mathrm{v}}(\vec{Q},\Omega) = &\frac{N^2g^2}{2\pi\beta v_F}\frac{\sqrt{\omega_0}}{64\pi^2}(\int_{0}^{\infty}\ud x\ud y-
\int_{0}^{\infty}\ud x\int_{-\infty}^{-\Omega}\ud y)\\
&\int\ud \bar{k}\ud \bar{k'}\frac{\bar{k}^2\bar{k'}^2-AB}{(\bar{k}^4+A^2)(\bar{k'}^4+B^2)((\bar{k}-\bar{k'})^2+2N^2\sqrt{|A-B|})}
\end{split}\ee
where \be\begin{split}
A &= \frac{2x+\Omega'}{8}\\
B &= \frac{2y+\Omega'}{8}
\end{split}\ee and \be\label{Omegap}
\Omega' = \frac{\Omega}{\omega_0},\ee
Rearranging the limits of integration in the second set of $x$, $y$ integrations, we obtain,
\be\begin{split}
\Pi_{\mathrm{v}}&(\vec{Q},\Omega) = \frac{N^2g^2}{2\pi\beta v_F}\frac{\sqrt{\omega_0}}{64\pi^2}\int_{0}^{\infty}\ud x\ud y
\int\ud \bar{k}\ud \bar{k'} \frac{\bar{k}^2\bar{k'}^2-AB}{(\bar{k}^4+A^2)(\bar{k'}^4+B^2)((\bar{k}-\bar{k'})^2+2N^2\sqrt{|A-B|})}\\
&-\frac{\bar{k}^2\bar{k'}^2+AB}{(\bar{k}^4+A^2)(\bar{k'}^4+B^2)((\bar{k}-\bar{k'})^2+2N^2\sqrt{|A+B|})}
\end{split}\ee
Performing the integration over $\bar{k}$ and $\bar{k'}$ we obtain, after some algebra,
\be\begin{split}\label{corr2}
\Pi_{\mathrm{v}}(\vec{Q},\Omega) &= \frac{N^2g^2\sqrt{\omega_0}}{2\pi\beta v_F}\frac{1}{64}\int_{0}^{\infty}\ud x\ud y\left(\frac{|\sqrt{A}-\sqrt{B}|}{\sqrt{AB}(4N^4(\sqrt{A}+\sqrt{B})+|\sqrt{A}-\sqrt{B}|^3)}\right.\\
& -\frac{A-B+2N^2\sqrt{|A-B|}}{2N\sqrt{B}\sqrt{\sqrt{|A-B|}}\left(\left(A-B-2N\sqrt{B}\sqrt{\sqrt{|A-B|}}\right)^2+4N^2\sqrt{|A-B|}
(N\sqrt{\sqrt{|A-B|}}+\sqrt{B})^2\right)}\\
& -\frac{B-A+2N^2\sqrt{|A-B|}}{2N\sqrt{A}\sqrt{\sqrt{|A-B|}}\left(\left(B-A-2N\sqrt{A}\sqrt{\sqrt{|A-B|}}\right)^2+4N^2\sqrt{|A-B|}(\sqrt{A}+N\sqrt{\sqrt{|A-B|}})^2\right)}\\
& -\frac{2(\sqrt{AB}+N^2\sqrt{A+B})}{\sqrt{AB}((A+B)^2+4N^4(A+B)+8N^2\sqrt{AB}\sqrt{A+B})}\\
& -\frac{A+B-2N^2\sqrt{A+B}}{2N\sqrt{B}\sqrt{\sqrt{A+B}}\left(\left(A+B+2N\sqrt{B}\sqrt{\sqrt{A+B}}\right)^2+4N^2\sqrt{A+B}(\sqrt{B}+N\sqrt{\sqrt{A+B}})^2\right)}\\
&\left. -\frac{A+B-2N^2\sqrt{A+B}}{2N\sqrt{A}\sqrt{\sqrt{A+B}}\left(\left(A+B+2N\sqrt{A}\sqrt{\sqrt{A+B}}\right)^2+4N^2\sqrt{A+B}(\sqrt{A}+N\sqrt{\sqrt{A+B}})^2\right)}\right)
\end{split}\ee\ewt We note that our result is symmetric w.r.t. to $A\leftrightarrow B$ as it should be, since the original 
integrals are symmetric w.r.t. $\omega \leftrightarrow \omega'$. 
Evaluating the remaining integrals, subtracting the $\Omega-$independent term coming from large $x,y$ (i.e., from high internal frequencies), and taking large $N$ limit, 
we find the leading term to be 
\be\label{pi2log}
\Pi_{\mathrm{v}}(\vec{Q},\Omega) = \frac{g^2\sqrt{\omega_0}}{2\pi\beta v_F}\frac{1}{2}\sqrt{\frac{\Omega'}{8}}\log\left(2N^2\sqrt{\frac{8}{\Omega'}}\right)
\ee
This is the result we quoted in the main text, Eq. (\ref{a_4}).

\end{document}